\documentclass[%
 reprint,
 superscriptaddress,
 amsmath,amssymb,
 aps,longbibliography
]{revtex4-2}

\usepackage{graphicx}
\usepackage{caption}
\usepackage{dcolumn}
\usepackage{bm}
\usepackage{gensymb}
\usepackage{physics}

\usepackage{hyperref}
\hypersetup{
    colorlinks,
    citecolor=black,
    filecolor=black,
    linkcolor=black,
    urlcolor=black
}

\preprint{APS/123-QED}

\begin{document}

\title{Ionic liquid gating of SrTiO$_3$ lamellas fabricated with a focused ion beam}

\author{Evgeny Mikheev}
\affiliation{Department of Physics, Stanford University, Stanford, CA, 94305, USA}
\affiliation{Stanford Institute for Materials and Energy Sciences, SLAC National Accelerator Laboratory, Menlo Park, CA 94025, USA}
\author{Tino Zimmerling}
\affiliation{Max-Planck-Institute for Chemical Physics of Solids, 01187 Dresden, Germany}
\author{Amelia Estry}
\affiliation{Laboratory of Quantum Materials (QMAT), Institute of Materials (IMX), École Polytechnique Fédérale de Lausanne (EPFL), Lausanne, Switzerland.}
\affiliation{Max-Planck-Institute for Chemical Physics of Solids, 01187 Dresden, Germany}
\author{Philip J. W. Moll}
\affiliation{Laboratory of Quantum Materials (QMAT), Institute of Materials (IMX), École Polytechnique Fédérale de Lausanne (EPFL), Lausanne, Switzerland.}
\affiliation{Max-Planck-Institute for Chemical Physics of Solids, 01187 Dresden, Germany}
\author{David Goldhaber-Gordon}
\affiliation{Department of Physics, Stanford University, Stanford, CA, 94305, USA}
\affiliation{Stanford Institute for Materials and Energy Sciences, SLAC National Accelerator Laboratory, Menlo Park, CA 94025, USA}

\begin{abstract}
In this work, we combine two previously-incompatible techniques for defining electronic devices: shaping three-dimensional crystals by focused ion beam (FIB), and two-dimensional electrostatic accumulation of charge carriers. The principal challenge for this integration is nanometer-scale surface damage inherent to any FIB-based fabrication. We address this by using a sacrificial protective layer to preserve a selected pristine surface. The test case presented here is accumulation of 2D carriers by ionic liquid gating at the surface of a micron-scale SrTiO$_3$ lamella. Preservation of surface quality is reflected in superconductivity of the accumulated carriers. This technique opens new avenues for realizing electrostatic charge tuning in materials that are not available as large or exfoliatable single crystals, and for patterning the geometry of the accumulated carriers.
\end{abstract}
\maketitle

\section*{Introduction}

Top-down fabrication using focused ion beam (FIB) milling has recently emerged as a powerful technique to explore physical properties of quantum materials \cite{Moll18,bruchhaus17}. Selective material removal by FIB enables definition of intricate device structures with features as small as hundreds of nanometers \cite{Moll16S,yasui20}, with explicit control of three-dimensional shape \cite{Moll16N,kim18} and orientation of conduction channel with respect to crystal axes \cite{Ronning17,Maniv21}.  Successful implementations of this technique have so far been in metallic compounds: high-$T_c$ superconductors \cite{Ooi02,yasui20}, high-conductivity delafossites \cite{Moll16S}, heavy fermion compounds \cite{Ronning17}, and arsenide semimetals \cite{Moll16N,Bachmann17,Ramshaw18}. The starting materials are bulk-grown single crystals, where top-down patterning by conventional lithography techniques is challenging due to chemical sensitivity and/or unavailability of high quality thin films.

At present, a major constraint on FIB-based patterning techniques is from surface damage \cite{Kato99,Giannuzzi99,Moll18}. Even very selective removal of material damages the entire exposed surface: typically a nanometer-scale surface layer is rendered partially amorphous and/or contains a high density of point defects. Though the thickness of the damage layer can be reduced by selection of ion species and acceleration voltages, it is rarely possible to reduce it below 2 nm \cite{kelley13}. For micron-scale structures with metallic bulk conduction such surface disorder is negligible, but for semiconductors it could screen the effect of electrostatic gates and dramatically-degrade mobility.

In this paper, we demonstrate voltage-driven accumulation of a superconducting 2D electron gas (2DEG) in a device defined by FIB. This combination is enabled by protecting a pristine surface from FIB damage with a sacrificial conductive layer that is removable in water. Electrons are subsequently accumulated at the preserved channel surface by ionic liquid gating. Here, this method is implemented on SrTiO$_3$, a material where electron doping and ionic liquid gating are both well known to drive a transition from an insulator to a two-dimensional superconducting metal \cite{Ueno08,sulpizio14,Pai18}. Observation of robust surface charge transport and superconductivity in our device demonstrates feasibility of integrating the FIB patterning toolbox with electrostatically tunable devices.

\begin{figure*}
\centering
\includegraphics[width=7in]{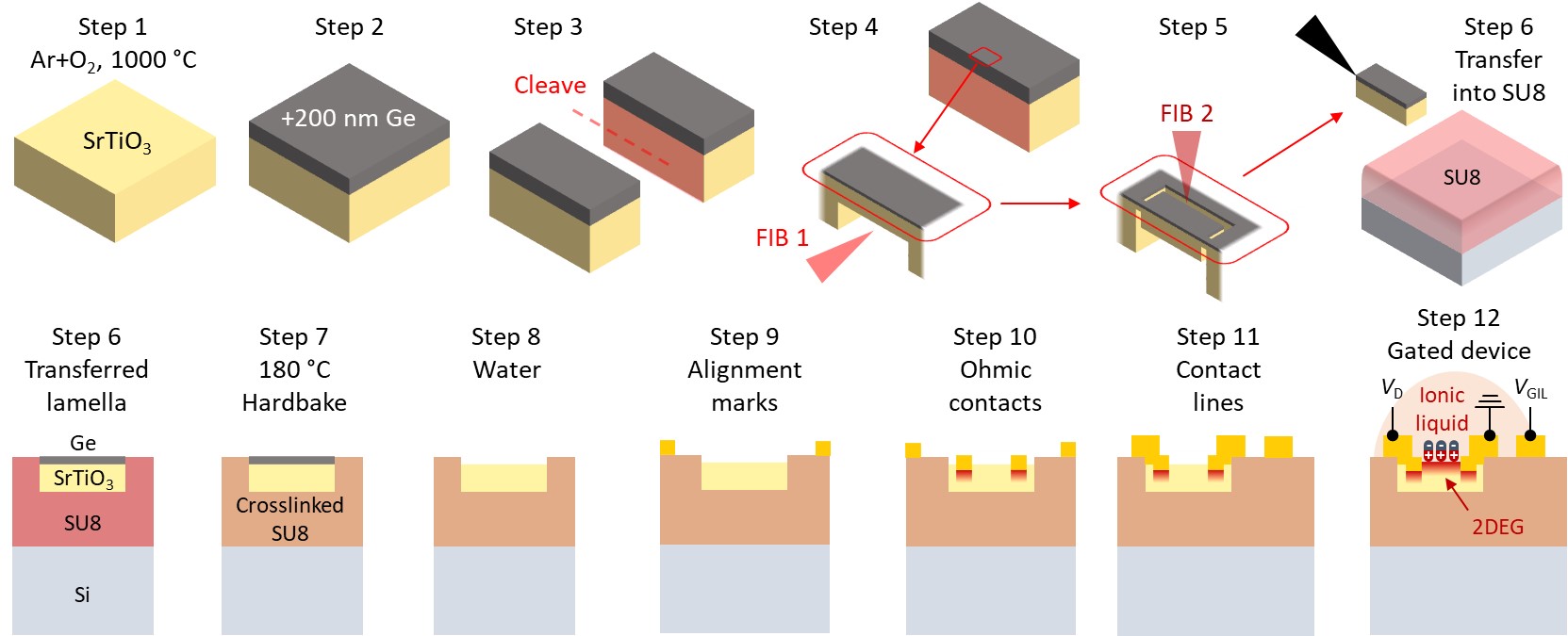}
\caption{\label{f1} Schematic illustration of the fabrication procedure steps, see text for details.}
\end{figure*}

\section*{Methods}

\begin{figure*}
\centering
\includegraphics[width=7in]{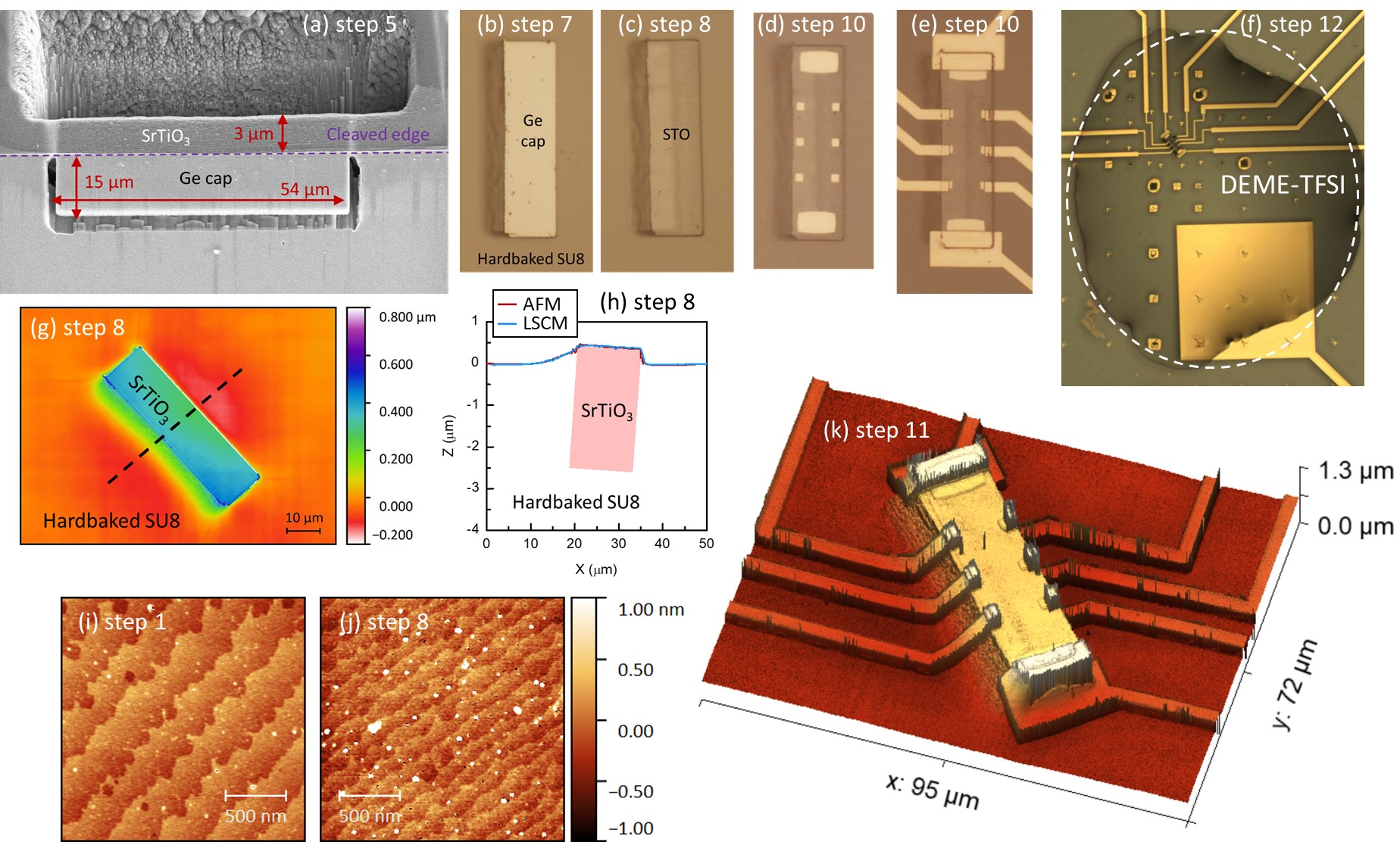}
\caption{\label{f2}(a) Scanning electron microscope image taken during FIB machining of the lamella. (b-f) Optical images of the devices after fabrication steps 7, 8, 10, 11, 12 (see text). (g) Height image taken with a 3D laser scanning confocal microscope (LSCM). (h) Comparison of height profile from LSCM and AFM. Both are taken near the middle of the lamella, as indicated by the dashed line in (g). (i) AFM image of the STO surface before sacrificial layer deposition, and (j) after its removal, on the surface of the lamella. (k) Perspective view of an LSCM height profile of the finished device.}
\end{figure*}

The fabrication flow described in this section aims to demonstrate that a first FIB-based patterning stage can be combined with subsequent conventional lithography techniques, ionic liquid gating, and cryogenic transport measurements. The account of each step of fabrication is accompanied by a schematic illustration (Fig.~\ref{f1}.) Fig.~\ref{f2} presents device images following selected fabrication steps.

\begin{itemize}
\item Step 1: A (001) Ti-terminated surface is prepared by soaking SrTiO$_3$ single-crystal substrates (MTI) for 20 minutes in de-ionized water heated to approximately 65 \textdegree C, and annealing for 2 hours at 1000 \degree C in Ar/O$_2$. 

\item Step 2: A sacrificial layer of 200 nm Ge is deposited by e-beam evaporation.

\item Step 3: The crystal is cleaved using a diamond scriber to create a smooth side surface.

\item Step 4: A rectangular lamella is defined by a Xenon FIB. The first cut is from the cleaved side of the crystal, defining the bottom surface of the lamella  (``FIB 1" in Fig.~\ref{f1}). 

\item Step 5: The second FIB cut is from the top, Ge-coated surface, defining the side walls of the lamella (``FIB 2'' in Fig.~\ref{f1}). Bridges on the sides are left to hold the sample in place as seen in Fig.~\ref{f2}a. Next, the FIB is used to polish the bottom surface of the lamella (same geometry as in the previous step, ``FIB 1'' in Fig.~\ref{f1}), creating a more uniform rectangular sample and allowing the surface of the Ge-coated STO to be flush with the SU8. For the final FIB process,  one of the bridges is cut and the other thinned down such  that the sample can be easily removed. 

\item Step 6: The lamella is broken out of the bulk crystal using an ex-situ micromanipulator. SU8 photoresist (GM1060, Gersteltec) is spun for 40 seconds at 4000 rpm on a 5x5 mm oxide-coated silicon chip; silicon is easily available, but other substrates would likely work similarly. With SU8 still in the viscous liquid state, the lamella is slowly transferred onto its surface.

\item Step 7: The chip is left in air for 10 hours, heated in air to 65 \textdegree C for 5 minutes, heated to 95 \textdegree C for 10 minutes, then cooled to room temperature. This process decreases the risk of the SU8 reflowing around the sample. Finally, to solidify the SU8, the chip is baked at 200 \textdegree C for 1 hour. All ramps are done at a rate of 2 \textdegree C per minute.  

\item Step 8: Removal of the sacrificial Ge layer. 5 minute exposure to an O$_2$ plasma for residue descum, followed by a 4 hour soak in de-ionized water heated to approximately 65 \textdegree C.

\item Step 9: First e-beam lithography (EBL) step. Deposition of alignment marks around the device via liftoff of 15/60 nm of Ti/Au. This and subsequent patterning steps are performed using spun-on PMMA resist,  a 30 kV EBL system, and a chilled 3:1 water:isopropyl alcohol mixture as a developer. Lift-off steps are performed with acetone followed by an isopropyl alcohol rinse.

\item Step 10: Second EBL step. Ohmic contact definition by ion milling (900 V beam voltage, 5 minutes) and lift-off of e-beam evaporated 10/40 nm of Ti/Au using the same patterned resist layer. A Hall bar-like contact geometry is defined, with two large source and drain contacts and six voltage probes.

\item Step 11: Third EBL step. Contact line, bond pads and side-gate deposition via liftoff of 10/200 nm of Ti/Au.

\item Step 12: The lamella device and the side gate are covered by an ionic liquid DEME-TFSI, and loaded into a dilution refrigerator system for charge transport characterization.
\end{itemize}

The overarching idea behind this process is to use hardbaked SU8 resist as a new planarized substrate for a micron-scale FIB-defined object that would otherwise be too small for conventional lithography techniques. After hardbaking, crosslinked SU8 is chemically robust to solvents, developers, and typical e-beam doses required for PMMA patterning. It is also sufficiently robust mechanically to withstand basic handling during device fabrication, and subsequent repeated thermal cycling to cryogenic temperatures.

One key enabler for this process is the resulting device surface topography, which is approximately flush between the FIB-defined lamella and the surrounding SU8. Remarkably, micron scale objects placed on the surface of liquid SU8 typically do not drown or float on their bottom surface, but instead partially submerge down to the exposed top surface. This is illustrated in Fig.~\ref{f1} in the device cross section for step 6 onwards. Images of actual device surface topography after hardbaking are shown in figures~\ref{f2}g and \ref{f2}h, using standard atomic force microscopy (AFM) and 3D laser scanning confocal microscopy (LSCM, Keyence VK-X250). The resulting boundary between SU8 and SrTiO$_3$ can easily be crossed with patterned 210 nm thick contact metal lines. In absence of the SU8 layer, metal lines would need to climb over a micron-scale cliff between the lamella top surface and the underlying solid substrate, which is challenging to realize robustly with reasonable metal thicknesses. Additionally, side wall contacts for some materials are likely to get shorted by conductive amorphous surface layers (see e.g. \cite{Bachmann17}).

A second key enabler is the use of a sacrificial Ge layer to preserve the SrTiO$_3$ surface from damage during the FIB milling steps. Ge satisfies two necessary criteria: 1) it is sufficiently robust to withstand secondary exposure to FIB (photoresist, in contrast, would be milled away too rapidly), 2) it can be selectively removed by simply soaking the device in water \cite{Melosh20}. The removal progress can be tracked optically (Fig.~\ref{f2}b,c) as the lamella surface color transitions from metallic (Ge) to transparent (SrTiO$_3$). The optical contrast on the device channel in Fig.~\ref{f2}c-e is from the lamella back side topography. We found that for devices exposed to FIB, heating the water above room temperature was necessary for complete removal, likely due to implantation of resputtered material by FIB into Ge. The AFM images in Fig.~\ref{f2}i,j present a first indication that the sacrificial layer is successful in shielding the intended device surface from FIB damage. The originally prepared SrTiO$_3$ surface has a terrace structure with unit cell steps resulting from substrate miscut angle. This morphology is successfully preserved throughout  Ge deposition, FIB milling and Ge removal. The difference seen in step spacing is a consequence of pre-existing miscut angle inhomogeneity across the 5x5 mm substrate.

For transport measurements, the electrical contact between gold bond pads and gold wire was made by cured conductive silver epoxy. The ionic liquid Diethylmethyl(2-methoxyethyl)ammonium bis(trifluoromethylsulfonyl)imide (DEME-TFSI) was deposited to cover the lamella and the adjacent side gate. The device was then loaded into a dilution refrigerator and vacuum pumped overnight to reduce ambient water contamination in the ionic liquid.

\section*{Results}

In the following, we present a detailed analysis of the superconducting transition in the FIB-prepared slabs. This is important for two reasons. First, the observed superconducting and normal state transport behavior compare well to prior measurements on ionic liquid gated SrTiO$_3$ surfaces in devices fabricated without FIB \cite{Ueno08,Lee11,Lee11b,Li12,Stanwyck13,Gallagher14,Gallagher15,Mikheev21} and to the closely-related 2DEG heterostructure systems SrTiO$_3$/LaAlO$_3$ and SrTiO$_3$/LaTiO$_3$ \cite{Reyren07,Caviglia10,Caprara11,Caprara13,Seri13,Biscaras14,Han14,Prawiroatmodjo16,Venditti19,Yin20}. These similarities leave no doubt that the main goal of pristine surface preservation was accomplished in this fabrication approach. Second, the SrTiO$_3$ sample dimensions (3$\times$15$\times$54~$\mu$m) are far smaller than in any previous SrTiO$_3$ device used for electrical transport studies. In this new regime, local strains fields can play an important role \cite{Bachmann19}. Moreover, these dimensions are comparable to the typical size of tetragonal domains, which spontaneously form in SrTiO$_3$ below a cubic-to-tetragonal transition near 100~K \cite{gray16} and are known to affect normal state \cite{kalisky13} and superconducting transport \cite{noad16,noad19}.

\begin{figure}
\centering
\includegraphics[width=3.5in]{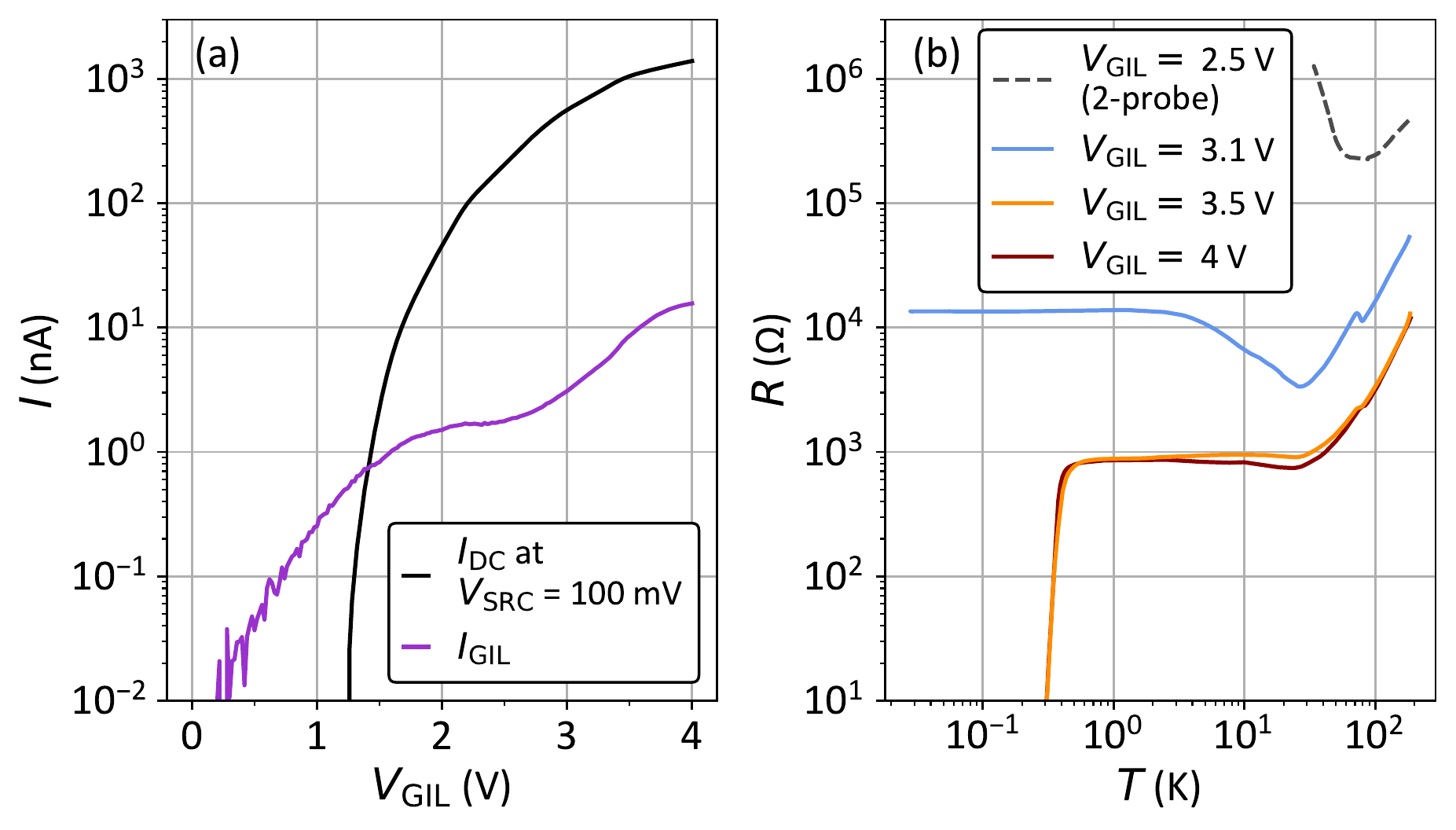}
\caption{\label{f3} Gate tuning of the charge transport in the lamella device channel. (a) Two-probe measurement near room temperature of DC current through the channel ($I_\text{DC}$) and leakage current to the gate electrode through the ionic liquid ($I_\text{GIL}$), as a function of gate voltage ($V_\text{GIL}$). (b) Temperature-dependent four probe measurements of device channel resistance, taken at different fixed values of $V_\text{GIL}$. For $V_\text{GIL}=$ 2.5~V, a two-probe measurement in the same configuration as (a) is shown.}
\end{figure}

The gate voltage $V_\text{GIL}$ is applied on the side gate near room temperature, polarizing the ionic liquid and accumulating a 2DEG on the surface of the lamella, as illustrated in step 12 of Fig.~\ref{f1}. Fig.~\ref{f3}a shows the 2-terminal measurement of DC current $I_\text{DC}$ through the device channel, measured while ramping $V_\text{GIL}$ up. A constant DC source bias $V_\text{SRC}=$ 100 mV is applied to one of the ohmic contacts, while all other contacts are grounded. Above $V_\text{GIL}=$ 1.2 V, carrier density accumulation in the channel results in a gradual transition from an insulator to a robust conductor. The device is then cooled down at constant $V_\text{GIL}$. The ionic liquid freezes near 220 K, fixing the 2DEG state until it is thermally cycled back to near room temperature for further adjustment of $V_\text{GIL}$. The temperature dependence of 4-terminal channel resistance was measured upon heating of the cryostat (Fig.~\ref{f3}b). For $V_\text{GIL}$ above 3~V, the channel becomes metallic. At $V_\text{GIL}=$ 3.1~V, a Kondo-like resistance upturn saturating at low temperature is seen, similarly to \cite{Lee11,Li12}. 

At $V_\text{GIL}=$ 3.5 and 4~V, robust metallicity is observed. The residual resistance ratio between 200~K and base temperature is 20, and $\approx 45$ if we extrapolate $R(T)$ to room temperature. These ratios imply low-temperature carrier mobility within the 100-1000~cm$^2$/Vs range typically observed in ionic liquid gated SrTiO$_3$ \cite{Ueno08,Lee11,Mikheev21}. The cusp feature near 80~K is typical for SrTiO$_3$-based 2DEGs measured upon heating, and is usually attributed to charge trapping and de-trapping at domain walls and/or point defects \cite{Seri13,Biscaras14,Yin20}.

A superconducting transition is observed in the mK range (Fig.~\ref{f3}b,c). Taking the midpoint of the resistance decrease as a measure of transition temperature $T_c$, it is 400 and 380 mK for $V_\text{GIL}=$ 3.5 and 4 V, respectively. 400 mK is the upper bound of previously reported $T_c$ in ionic liquid gated SrTiO$_3$ \cite{Ueno08,Lee11b,Stanwyck13,Gallagher14,Gallagher15,Mikheev21}. It is also notably higher than any such devices previously fabricated in the authors' laboratory (without FIB), which typically have $T_c < $ 350~mK \cite{Stanwyck13,Gallagher14,Gallagher15,Mikheev21}. Superconductivity in SrTiO$_3$ is sensitive to strain, likely by tuning proximity to ferroelectric quantum criticality \cite{Ahadi19,Herrera19,enderlein20}. The unexpectedly high $T_c$ in a micron-scale lamella device may thus reflect local strain fields from differential thermal expansion with the surrounding SU8 and the underlying silicon substrate \cite{Bachmann19}.

\begin{figure*}
\centering
\includegraphics[width=7in]{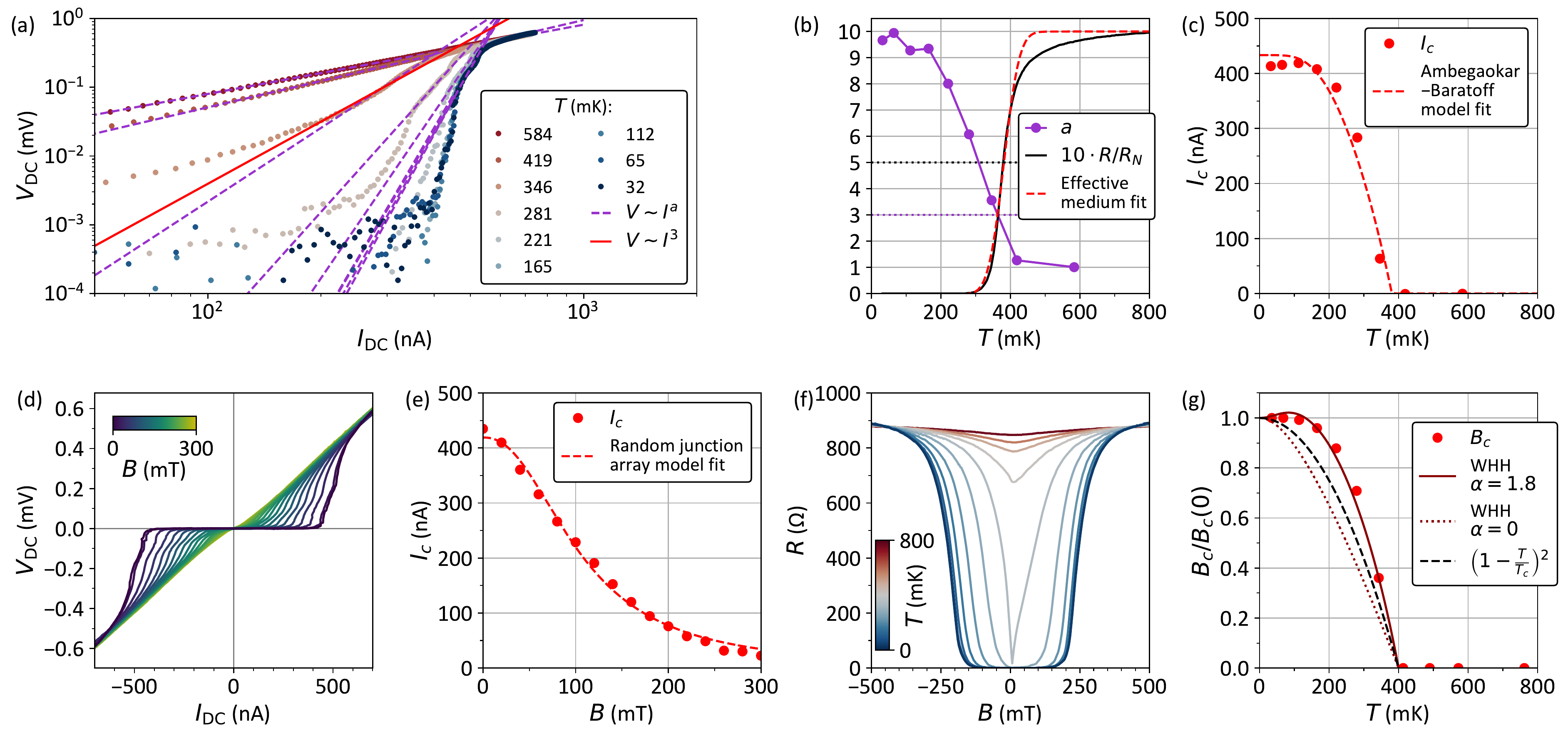}
\caption{\label{f4}Suppression of superconductivity with DC current $I_\text{DC}$, temperature $T$, and perpendicular field $B$. (a) 4 probe DC measurement of current-voltage non-linearity in the device channel. Purple dashed lines are power law fits, with the red solid line indicating $a=3$ exponent criterion for a BKT transition. (b) $T$ dependence of the exponent $a$ from power law fits in (a). Solid line is the $T$ dependence of channel resistance (from Fig.~\ref{f3}b), scaled by a factor of 10 for comparison with $a(T)$. Dotted lines show the transition point criteria $a=3$ and $R/R_N = 0.5$. Dashed red line is a fit to equation~\ref{eqEM}. (c) $T$ dependence of the critical current and a fit to equation~\ref{eqAB}. (d) Same measurement as in (a), but as a function of $B$ at 32 mK. (e) Critical current in (d) and a fit to equation~\ref{eqIcB}. (f) 4 probe AC resistance measurement as a function of $B$ and $T$. (g) $T$ dependence of the critical field, normalized to base temperature value. Black dotted line is the 3D BCS lineshape. Red dotted and solid lines are WHH model lineshapes \cite{Werthamer66}, with $\alpha=$ 0 and 1.8. $V_\text{GIL}=$ 4~V in (a-e), 3.5~V in (f, g).}
\end{figure*}

Fig.~\ref{f4} illustrates the critical scales for suppression of superconductivity with DC current $I_\text{DC}$ and magnetic field $B$.
Fig.~\ref{f4}a shows a strong non-linearity in 4-terminal DC measurements. $I_\text{DC}$ is the DC current excitation sourced through the large source and drain contacts, and $V_\text{DC}$ the DC bias measured by adjacent voltage probes on the channel. This non-linearity follows a power law $I_\text{DC} \sim (V_\text{DC})^{a}$ with a temperature-dependent exponent $a(T)$ shown in Fig.~\ref{f4}b. This is reminiscent of the BKT Berezinskii–Kosterlitz–Thouless (BKT) transition framework, based on breaking of vortex-antivortex pairs by DC current in a 2D superconductor. The BKT transition point signaled by $a=3$ is at 365 mK, only slightly below $T_c=$ 380 mK. Very similar transitions have been reported in LaAlO$_3$/SrTiO$_3$ 2DEGs \cite{Reyren07,Han14,Monteiro17,Venditti19}. As pointed out in \cite{Monteiro17,Venditti19} a failure point of the BKT description is that it entails a discontinuous jump in $a$, rather than a gradual transition. The rate of increase in $a$ with $T$ in Fig.~\ref{f4}b is similar to that in \cite{Monteiro17,Venditti19}.

An alternate picture of this transition is in terms of mesoscopically inhomogeneous superconductivity. A common approach for SrTiO$_3$-based 2DEGs is to describe the system by effective medium theory \cite{Venditti19,Caprara11,Caprara13}: normal state regions coexist with superconducting puddles with a probabilistic distribution of $T_c$. For the thermally-driven resistive transition this predicts
\begin{equation}
\label{eqEM}
R (T)=R_N \left(1-w+w\,\mathrm{ erf}\left(\frac{T-\overline{T}_c}{\sqrt{2}\gamma}\right)\right),
\end{equation}
Where $w$ is the fraction of superconducting puddles,  $\overline{T}_c$ and $\gamma$ are the average value and width of the Gaussian distribution for $T_c$, $R_N$ is the normal state resistance. A fit line with $w=0.5$, $\overline{T}_c=$ 380 K, $\gamma=$ 35 K is shown in Fig.~\ref{f4}b. Similarly to \cite{Caprara13,Prawiroatmodjo16,Venditti19}, this simple model successfully accounts for the large width of the resistance drop, but fails to describe the prolonged tail above $T_c$. The latter shortcoming may reflect a more complex statistical distribution of local $T_c$, for example a bimodal distribution including a second set of puddles with $T_c\approx$ 400-700~mK~\cite{Caprara11}.

Further support for the inhomogeneous superconductivity picture comes from analysis of the superconducting critical current $I_c$ (Fig.~\ref{f4}c-e). $I_c$ was defined as $I_\text{DC}$ at which $V_\text{DC}$ reaches 5~$\mu$V. Similarly to  \cite{Prawiroatmodjo16,Venditti19}, its temperature dependence is well described by the Ambegaokar-Baratoff model of a dirty superconducting weak link:
\begin{equation}
\label{eqAB}
I_c (T)=\frac{\pi\Delta (T)}{2eR_{N}^{*}}\tanh \frac{\Delta (T)}{2k_B T},
\end{equation}
Where $\Delta (T)=\Delta (0)\tanh{(1.74\sqrt{(T_c-T)/T})}$ is the BCS superconducting gap, and $R_N^*$ is the normal state weak link resistance. For the inhmogeneous case, this model corresponds to an array of Josephson junctions between superconducting puddles with the same functional form of $I_c (T)$ and a total normal state resistance $R_N^*$ \cite{Prawiroatmodjo16}. The fit in Fig.~\ref{f4}c is to a normalized lineshape with $I_c(0)=$ 430 nA and $T_c=$ 380 mK.


At base temperature, the suppression of $I_c$ with a magnetic field $B$ orthogonal to the 2DEG plane (Fig.~\ref{f4}d) is well described by the random junction array model \cite{Prawiroatmodjo16,muller92,vanderLaan01}:
\begin{equation}
\label{eqIcB}
I_c (B)=I_c(0) \cdot\frac{1}{1+(B/B_0)^\beta},
\end{equation}
Where $B_0=\Phi_0/\pi A_0$, $A_0$ is the junction area, $\beta$ is an array geometry-dependent exponent, typically 1-2 \cite{Prawiroatmodjo16,muller92,vanderLaan01}. Fitting in Fig.~\ref{f4}e gives $\beta = 2.3$, $\sqrt{A_0} = 80$ nm, consistent with a mesoscopic array of Josephson junctions.

Figures~\ref{f4}f,g show how superconductivity is suppressed with $B$ and $T$ in a 4-probe AC measurement at zero DC bias (for $V_\text{GIL}=$ 3.5~V). The temperature dependence of the critical field $B_c$ (defined as the midpoint of the resistance drop in $B$) presents a very rapid decrease close to $T_c$. The concave curvature is stronger than both the 2D BCS dependence ($B_c/B_c(0)=1-T/T_c$) and 3D BCS dependence ($B_c/B_c(0)=\sqrt{1-T/T_c}$). A similarly strong curvature in $B_c(T)$ can appear in the Werthamer-Helfand-Hohenberg (WHH) model \cite{Werthamer66,kim12,solenov17} in the limit of strong orbital depairing ($\alpha$) and low spin-orbit scattering strength ($\lambda_\text{SO}$) \cite{solenov17}. The data in Fig.~\ref{f4}g are well matched by the normalized WHH curve with $\alpha=1.6$ and $\lambda_\text{SO}=0$. Spin-orbit scattering values in STO span a wide range as a function of carrier density and between different experiments \cite{Pai18}. So negligible $\lambda_\text{SO}$ is not too surprising, but it is a departure from the common pattern of high spin-orbit coupling approximately coinciding with optimal superconducting $T_c$ \cite{Caviglia10,Pai18}.

Doping-induced insulator-to-metal transition and inhomogeneous superconductivity are key features of the carrier density-temperature phase diagram in SrTiO$_3$-based 2DEGs. Both are preserved in our device despite its exposure to FIB milling. These are transport-based indications that our approach can preserve a chosen channel surface and the relevant host material physics.

\section*{Conclusions}

We have demonstrated a unique combination of top-down patterning with FIB, EBL, and voltage-driven accumulation of carrier density. The key enabler is the use of sacrificial Ge layer to protect a pristine surface from FIB damage. This technique can be used to fabricate gated devices from materials other than SrTiO$_3$, for which device fabrication by standard lithography is not feasible because large single crystal or thin films are not available.

Separately, combining ionic liquid gating and/or conventional electrostatic gating with three dimensional shape control on the micron scale by FIB presents exciting opportunities. For example, 2DEG functionality could be controlled with engineered local strain \cite{Bachmann19} or piezoelectric actuation of stretching, compression, or bending of FIB-defined lamellas \cite{kim18}. Provided the lamella material is piezoelectric or has a large electrostrictive coefficient (as is the case for SrTiO$_3$ \cite{Grupp97}) bulk or surface acoustic waves could be generated and acoustoelectrically coupled to the induced 2DEG \cite{uzun20}. Here, FIB would both define key dimensions and thus frequencies of resonant acoustic waves, and enable releasing micron-scale structures from a larger substrate.

\renewcommand{\bibsection}{\section*{References}}
\bibliography{references.bib}

\vspace{3mm}

\section*{Acknowledgments}

We thank Carsten Putzke for help with focused ion beam work; Trevor Petach, Xiao Zhang, and Joe Finney for help with the development of lithographic fabrication processes; and Ilan Rosen and Eli Fox for help with cryogenic measurements. Our use of Germanium as a water-soluble protective layer was inspired by discussions with Nicholas Melosh, prior to publication of his work on Germanium as a protective layer for lithography on organic materials.

\textbf{Funding:} Experimental work (fabrication and measurement) by E.M. was primarily supported by the Air Force Office of Scientific Research through grant no. FA9550-16-1-0126. E. M. was also supported by the Nano- and Quantum Science and Engineering Postdoctoral Fellowship at Stanford University and (during paper-writing and analysis) by the Department of Energy,  Office of Science, Basic Energy Sciences, Materials Sciences and Engineering Division, under Contract DE-AC02-76SF00515. Measurement infrastructure was funded in part by the Gordon and Betty Moore Foundation’s EPiQS Initiative through grant GBMF3429 and grant GBMF9460. Part of this work was performed at the Stanford Nano Shared Facilities (SNSF), supported by the National Science Foundation under award ECCS-1542152.

\textbf{Author contributions:} E.M., P.J.W.M., T.Z. and D.G.-G. designed the experiment. T.Z., A.E. and P.J.W.M. carried out FIB-based device fabrication. E.M. carried out EBL-based device fabrication and cryogenic measurements. All authors discussed the results. E.M. wrote the manuscript with input from other authors.

\end{document}